\documentclass[twocolumn,fleqn]{autart}    

\usepackage{graphicx}          
\usepackage{amsmath,amsfonts}
\usepackage{algorithm}
\usepackage{algorithmicx}
\usepackage{algpseudocode}
\usepackage{array}
\usepackage{textcomp}
\usepackage{stfloats}
\usepackage{url}
\usepackage{verbatim}
\usepackage{amssymb}
\usepackage{color}
\usepackage{epstopdf}
\usepackage[numbers,sort&compress]{natbib}
\usepackage{multirow}
\usepackage{mathrsfs}
\usepackage{enumerate}
\usepackage{bm}
\usepackage{parskip}
\usepackage{setspace}
\usepackage{nowidow}
\usepackage{indentfirst}
\begin{document}
\begin{frontmatter}

\title{Inverse Optimal Control for Linear Quadratic Tracking with Unknown Target States\thanksref{footnoteinfo}} 
\thanks[footnoteinfo]{This work was supported by Chongqing Natural Science Foundation CSTB2023NSCQ-JQX0018, National Natural Science Foundation of China (NSFC) (Grant No. 62473046). The material in this paper was not presented at any conference. }
\thanks{Corresponding author C.~Yu.}

\author[CIC,BIT]{Yao Li}\ead{yao.li@bit.edu.cn},    
\author[CIC,BIT]{Chengpu Yu}\ead{yuchengpu@bit.edu.cn},               
\author[BIT]{Hao Fang}\ead{fangh@bit.edu.cn},  
\author[BIT,HIT]{Jie Chen}\ead{chenjie@bit.edu.cn}

\address[CIC]{Beijing Institute of Technology Chongqing Innovation Center, Chongqing, 401120, China} 
\address[BIT]{School of Automation, Beijing Institute of Technology, Beijing, 100081, China}     	 
\address[HIT]{Harbin Institute of Technology, Harbin 150001, China}                   			 


\begin{abstract}                          
This paper addresses the inverse optimal control for the linear quadratic tracking problem with a fixed but unknown target state, which aims to estimate the possible triplets comprising the target state, the state weight matrix, and the input weight matrix from observed optimal control input and the corresponding state trajectories. Sufficient conditions have been provided for the unique determination of both the linear quadratic cost function as well as the target state. A computationally efficient and numerically reliable parameter identification algorithm is proposed by equating optimal control strategies with a system of linear equations, and the associated relative error upper bound is derived in terms of data volume and signal-to-noise ratio (SNR). Moreover, the proposed inverse optimal control algorithm is applied for the joint cluster coordination and intent identification of a multi-agent system. By incorporating the structural constraint of the Laplace matrix, the relative error upper bound can be reduced accordingly. Finally, the algorithm's efficiency and accuracy are validated by a vehicle-on-a-lever example and a multi-agent formation control example.
\end{abstract}
\end{frontmatter}

\section{Introduction}
Inverse Optimal Control (IOC), pioneered by \citep{bx1}, identifies the underlying reward or penalty mechanism from observed optimal control trajectories. It resembles inverse reinforcement learning (IRL), which assumes a parametrized cost function identified via optimal input and state trajectories. The distinction between IOC and IRL is often unclear; some view IRL as IOC with (partially) unknown dynamics \citep{bx30}. The authors suggest that IOC applies when the identified cost function is used for optimal control, while IRL applies in reinforcement learning. IOC facilitates knowledge transfer to analogous scenarios with new dynamics. Its applications include robotics \citep{b11, a9, b21, bx2, bx3} and other domains \citep{a10, bx4, bx9}.

IOC problems are classified into finite-horizon scenarios, yielding time-variant strategies, and infinite-horizon scenarios, yielding time-invariant ones. Effective methods exist for infinite-horizon IOC \citep{bx14, bx20, bx21, bx19}. \citep{bx12} explores IOC for averaged-cost linear quadratic regulators, accounting for process and observation noises. For finite-horizon IOC problems, a few approaches based on open-loop information structures, such as Pontryagin’s maximum principle, have been proposed in \citep{bx27, bx22, bx13}. Despite the advancements, many identification methods neglect process or observation noises \citep{d5,b12,a5,bx23,a9,d3,bx7,bx6,b1,d7}. With only observation noise, \citep{b21} demonstrates consistent results as the number of trajectories tends to infinity. IRL developments, like entropy maximization \citep{n1}, incorporate process noise but face challenges in characterizing trajectory distributions. \citep{n3} applies an approach to various systems, providing consistency guarantees for linear systems with zero-mean additive noise, polytopic constraints, and quadratic objectives. However, finite-horizon IOC methods encounter difficulties with both process and observation noises \citep{d8}. While \citep{bz5} handles both noises, it assumes known statistical properties of them; simultaneously addressing both types of noises with less prior information remains an open problem.

The simplicity and broad applicability of linear quadratic optimal control have resulted in a multitude of related IOC applications \citep{bx15, bx16, bx17, bx18}.  
This, also in the light of IRL, which in the absence of inequality constraints, allows for consideration of process noise by construction (considering the state transitions as a Markov decision process) \citep{n2}.
In recent years, IOC methods have been developed for linear quadratic tracking (LQT) optimal control, with research on IOC for LQT in continuous-time infinite-horizon scenarios conducted in \citep{bx28, bx30}. Additionally, \citep{bx29} presents a data-driven IOC method for LQT in discrete-time infinite-horizon settings, which does not require system dynamics information. Finite-horizon LQT-IOC problems for state feedback control and output feedback control are explored in \citep{b21} and \citep{bz1}, respectively. Challenges arise when the target state is unknown, given the coupling between the target state and the state weight matrix in the LQT cost function.

This paper focuses on solving the IOC for LQT within the discrete-time finite-horizon scenario, aiming to jointly determine the state weight matrix,  the input weight matrix, and the target state. The concerned IOC problem is tackled by equivalently transforming the optimal control policy into linear equations. Addressing the coupling term in the LQT cost function involves initially determining the state weight matrix and input weight matrix, followed by estimating the target state from a specific solution space. When parameters possess structural prior information, projection techniques can be used to minimize errors effectively. For the optimal formation control problem, the Laplacian matrix generates state weight matrices with duplicate and zero elements, which can be projected through (half-)vectorization operations, to maintain linearity in solving the IOC problem.

The contributions of this paper are summarized as follows:

(i) {An efficient IOC method is proposed for LQT problems with both process and observation noise. Unlike existing approaches such as \citep{bz5}, it avoids the need for noise covariance information and identifies the target state $d$, the state cost matrix $Q$, and the input cost matrix $R$ jointly. Sufficient identifiability conditions are provided under mild constraints and a relative error bound is established with a provable convergence rate of $\mathcal{O}(M^{-0.5})$, where $M$ is the number of observed trajectories.}

(ii) The specialization of the proposed method is demonstrated in the IOC of the multi-agent formation control with unknown fixed interconnection topology. It is theoretically shown that integrating structural prior knowledge can reduce algorithmic errors, illustrating the extensibility of the proposed method in high-dimensional systems with certain structural information. 

\emph{Notations:}
$I_{n}$ is the $n\times n$ identity matrix; $\bm{0}_{n\times p}$ is the zero matrix (subscripts omitted if unambiguous).
$\mathbf{1}_{N}$ is an $N$-dimensional all-ones vector.
$\mathbb{R}^n$ is the $n$-dimensional real linear space. $\mathbb{S}^n$, $\mathbb{S}_+^n$, and $\mathbb{S}_{++}^n$ denote sets of $n\times n$ symmetric, positive semi-definite, and positive definite matrices, respectively.
$\otimes$ is the Kronecker product.
For $A\in\mathbb{R}^{n\times n}$, $vec(A)$ and $vech(A)$ are the $n^2\times1$ and $n(n+1)/2\times1$ vectors stacking $A$'s columns and lower triangular part, respectively. For $A\in\mathbb{S}^n$, the duplication matrix $\mathcal{D}_n$ satisfies $\mathcal{D}_n vech(A)=vec(A)$.
$A^\dagger$ is the pseudoinverse; $\mathcal{P}^{\text{ker}(A)}=I-A^\dagger A$ projects onto the kernel space $\text{ker}(A)$. $A^{-1}$ is the inverse for invertible $A$. $A^\top$ is the transpose; $\|\cdot\|$ is the $L_2$ norm.
For full-column-rank $A$, $\text{cond}(A)$ is its $L_2$-norm condition number, the ratio of largest to smallest singular values.
$\mathbb{E}$ denotes expectation.
$\sigma_{\text{min}}(\cdot)$ denotes the smallest singular value of a matrix.
$\text{diag}({\phi})$ is the (block) diagonal matrix with $\phi$ on its diagonal.

\section{Problem Formulation}\label{sec2}

The discrete-time finite-horizon LQT problem with process noise is considered:
\begin{subequations}\label{fm1}
\begin{align}
\min_{\substack{u_k}}~&\mathbb{E}\left\{\sum_{k=0}^{K} e_{k+1}^\top Q e_{k+1} + u_k^\top R u_k \right\} \label{fm1a}\\
\text{s.t.}~&x_{k+1} = A x_k + B u_k + w_k \label{fm1b}\\
&e_{k+1} = x_{k+1} - d \label{fm1c}, \quad k = 0,1,\dots,K
\end{align}
\end{subequations}
where \( x_k \in \mathbb{R}^n \), \( u_k \in \mathbb{R}^p \), and \( d \in \mathbb{R}^n \). The process noise \( w_k \) is assumed to be white and zero-mean, uncorrelated with the initial state \( x_0 \). The matrices \( A \), \( B \), \( Q \), \( R \) are conformable, and \( K \) is a finite control horizon.

\begin{assum}\label{as1}
The weight matrices satisfy \( Q \in \mathbb{S}_+^n \), \( R \in \mathbb{S}_{++}^p \).
\end{assum}

\begin{lem}[\citep{bertsekas2012dynamic}]\label{lemma1}
Under Assumption \ref{as1}, the optimal control input for problem \eqref{fm1} is given by
\begin{align}
u_k = \pi_k(Q,R,d,x_k)\label{fm2}
\end{align}
where
\begin{subequations}\label{fm3}
\begin{align}
\pi_k(Q,R,d,x_k) &= \begin{bmatrix} \Psi_k(Q,R) & \psi_k(Q,R,d) \end{bmatrix} \begin{bmatrix} x_k \\ 1 \end{bmatrix} \label{fm3a}\\
\Psi_k(Q,R) &= -\left(R + B^\top P_{k+1} B\right)^{-1} B^\top P_{k+1} A \label{fm3b}\\
\psi_k(Q,R,d) &= -\left(R + B^\top P_{k+1} B\right)^{-1} B^\top \eta_{k+1} \label{fm3c}
\end{align}
\end{subequations}
and  \( P_k \), \( \eta_k \) are determined by the following Riccati equations for $k=1,2,\cdots,K$
\begin{subequations}\label{fm4}
\begin{align}
\eta_k &= q + A^\top (P_{k+1} B \psi_k + \eta_{k+1}) \label{fm4a}\\
P_k &= Q + A^\top (P_{k+1} B \Psi_k + P_{k+1} A) \label{fm4b}
\end{align}
\end{subequations}
with terminal conditions \( P_{K+1} = Q \), \( \eta_{K+1} = q := -Qd \). Furthermore, \( P_{k+1} \in \mathbb{S}_+^n \) for all \( k = 0,\dots,K \).
\end{lem}

\begin{rem}
Although the state transition \eqref{fm1b} includes process noise, it does not appear in the optimal control policy \eqref{fm2}. Moreover, the target state \( d \) and the matrix \( Q \) are decoupled in the policy formulation, which is beneficial for inverse identification.
\end{rem}

\begin{assum}\label{as2}
The matrix pair \( (A,B) \) is known and controllable; additionally, \( A \) is invertible and \( B \) has full column rank.
\end{assum}

\begin{assum}\label{as3}
The control horizon \( K \ge 2n \).
\end{assum}

These assumptions are commonly reasonable in discretized continuous-time systems. For instance, if \( A = e^{\mathcal{A}\tau} \), where \( \mathcal{A} \) is the continuous-time system matrix and \( \tau \) is the sampling interval, then \( A \) is invertible. The full column rank of \( B \) ensures unique input-to-state mappings. These assumptions are also required in Section~\ref{sec3} for identifiability, as supported by \citep{d3}, \citep{b1}, and \citep{d8}.

\begin{prob}\label{prob1}
Given \( M \) optimal input and state trajectories of length \( K \), denoted by \( \{x_{k,i}, u_{k,i}\}_{k=0,i=1}^{K,M} \), the IOC problem aims to determine the set of triplets \( (Q, R, d) \) such that
\(u_{k,i} = \pi_k(Q,R,d,x_{k,i})\)
,
under Assumptions \ref{as1}, \ref{as2} and \ref{as3}.
\end{prob}

\begin{rem}
The observed inputs \( u_{k,i} \) are required to conform to the policy \( \pi_k \), even if the trajectories do not exactly minimize \eqref{fm1}. The following result still applies to approximately optimal trajectories, provided that the control law aligns with the fixed cost structure.
\end{rem}

\section{Identifiability analysis}\label{sec3}
Identifiability analysis for the IOC problem is addressed in this section. Process noise effects are considered, while observation noise is examined in Section \ref{sec4b}. 

Given data $\{x_{k,i},u_{k,i}\}_{k=0,i=1}^{K,M}$ adhering to (\ref{fm2}), since $R+B^\top P_{k+1}B$ is invertible by Lemma \ref{lemma1}, if the following matrix $$X_k=\begin{bmatrix}x_{k,1}&\cdots&x_{k,M}\\1&\cdots&1\end{bmatrix}$$has full row rank, then the equations \eqref{tt1}-\eqref{tt0e} below is equivalent to \eqref{fm2}-\eqref{fm4}  under Assumption \ref{as1}.
\begin{subequations}\label{tt0}
\begin{align}
&B^\top \begin{bmatrix}{P}_{k+1}&{\eta}_{k+1}\end{bmatrix}Y_{k}+{R}U_{k}=\bm{0}\label{tt1}\\
&A^\top \begin{bmatrix}{P}_{k+1}&{\eta}_{k+1}\end{bmatrix}Y_{k}+\begin{bmatrix}{Q}-{P}_k & {q}-{\eta}_k\end{bmatrix}X_k=\bm{0}\label{tt0b}\\
&\begin{bmatrix}{P}_{K+1}&{\eta}_{K+1}\end{bmatrix}=\begin{bmatrix}{Q}&{q}\end{bmatrix}\label{tt0c}\\
&Y_{k} = \begin{bmatrix}A & \bm{0} & B \\ \bm{0} & 1 & \bm{0} \end{bmatrix}\begin{bmatrix}X_{k} \\ U_{k}\end{bmatrix}\label{tt2}\\
&U_{k} = \begin{bmatrix}u_{k,1} & \cdots & u_{k,M}\end{bmatrix} \label{tt0e}\\
&{P}_{k}\in \mathbb{S}^n, {R}\in \mathbb{S}^p\setminus\{\bm{0}\}\label{tt0f}
\end{align}
\end{subequations}
where \eqref{tt0f} is a necessary condition for Assumption \ref{as1}. The equivalence above 
can be verified by substituting \eqref{tt2}-\eqref{tt0e} into \eqref{tt1}-\eqref{tt0b} with an intermediate step $U_{k} = \begin{bmatrix}
    \Psi_k&\psi_k
\end{bmatrix}X_{k}$, and it
implies that the true parameters $\left({Q},{R},{q},\{{P}_k\}_{k=1}^{K+1},\{{\eta}_k\}_{k=1}^{K+1}\right)$ satisfy \eqref{tt0} under Assumption \ref{as1}. Namely, the solution to \eqref{tt0} exists. 
On the other hand, for given $x_k$ and any $\alpha>0,\lambda\in\mathbb{R}^n$, the triplet $(Q,R,d)$ produces identical $u_k$ as $(\alpha Q,\alpha R,d+\mathcal{P}^{\text{ker}(Q)}\lambda)$, owing to cost (\ref{fm1a}), where $\mathcal{P}^{\text{ker}(Q)}=I-Q^\dagger Q$.  
Thus, such parameters solve the IOC problem.  
The following theorem states sufficient identifiability conditions.

\begin{thm}\label{thm1}
Consider Problem \ref{prob1} with unknown $(Q,R,d)$, assume that the following conditions are satisfied:
(i) Assumptions \ref{as1}, \ref{as2} and \ref{as3} hold;
(ii) $\mathcal{Q}=BR^{-1}B^\top+Q\in\mathbb{S}_{++}^n$ and $\mathcal{Q}A\ne A\mathcal{Q}$;
(iii) For $k=0,\dots,K$, $X_k$ has full row rank.  
Then, for any solution $\left(\hat{Q},\hat{R},\hat{q},\{\hat{P}_k\}_{k=1}^{K+1},\{\hat{\eta}_k\}_{k=1}^{K+1}\right)$ to \eqref{tt0}, there exists a nonzero scalar $\alpha$ such that $\hat{Q}=\alpha Q,\hat{R}=\alpha R,\hat{q}=\alpha q$.  
Furthermore, if $Q$ and $d$ comply with $\mathcal{P}^{\text{ker}(Q)}d=\bm{0}$, then $\hat{d}:=-\hat{Q}^\dagger\hat{q}= d$.
\end{thm}

\begin{pf}
To begin with, the strict conditions are temporarily assumed that both $\hat{R}$ and $\hat{R}+B^\top \hat{P}_{k+1}B$ are either positive or negative definite. At the end of the proof, it will be shown that the strict conditions are indeed needless.
Given data satisfying (\ref{fm2}), the equations \eqref{fm3a} and \eqref{tt0e} imply $U_k=\begin{bmatrix}G_k&g_k\end{bmatrix}X_k$ for some $G_k,g_k$. Substituting it into \eqref{tt1}-\eqref{tt2} for solution $\left(\hat{Q},\hat{R},\hat{d},\{\hat{P}_k\},\{\hat{\eta}_k\}\right)$ yields that
\begin{small}
\begin{subequations}
\begin{align}
&\left\{A^\top \hat{P}_{k+1}B\begin{bmatrix}G_k&g_k\end{bmatrix}+\right.\notag\\
&\quad\left.\begin{bmatrix}\hat{Q}+A^\top \hat{P}_{k+1}A-\hat{P}_{k}&\hat{q}+A^\top\hat{\eta}_{k+1}-\hat{\eta}_k\end{bmatrix}\right\}X_k=\bm{0}\label{fm8b}\\
&\left\{(\hat{R}+B^\top\hat{ P}_{k+1}B)\begin{bmatrix}G_k&g_k\end{bmatrix}+\begin{bmatrix}B^\top \hat{P}_{k+1}A&B^\top\hat{\eta}_{k+1}\end{bmatrix}\right\}X_k=\bm{0}
\label{fm8a}
\end{align}
\end{subequations} 
\end{small}Then the full row rank of $X_k$ implies that 
\begin{subequations}\label{ntt2}
\begin{align}
\hat{\eta}_k&=\hat{q}+A^\top(\hat{P}_{k+1}Bg_k+\hat{\eta}_{k+1})\label{tt6}\\
\hat{P}_k&= \hat{Q}+A^\top(\hat{P}_{k+1}BG_k+ \hat{P}_{k+1}A) \label{tt7}
\end{align}
\end{subequations}
from \eqref{fm8b}. With $\hat{R}+B^\top \hat{P}_{k+1}B$ either positive or negative definite, \eqref{fm8a} gives that 
$G_k=-(\hat{R}+B^\top \hat{P}_{k+1}B)^{-1}B^\top \hat{P}_{k+1}A$, as well as $g_k=-(\hat{R}+B^\top \hat{P}_{k+1}B)^{-1}B^\top\hat{\eta}_{k+1}$.
Thus $\pi_k(\hat{Q},\hat{R}, \hat{d}, x_{k,i})=u_{k,i}=\pi_k(Q, R, d, x_{k,i})$ is derived by \eqref{fm2}-\eqref{fm4}. Hence, condition (iii) eliminates $X_k$ and
\begin{align}
\Psi_k(Q, R)&=\Psi_k(\hat{Q},\hat{R})\label{t1}\\
\psi_k(Q, R, d)&=\psi_k(\hat{Q},\hat{R}, \hat{d})\label{t2}
\end{align}
Left-multiplying \eqref{t1} by $B$ and adding $A$ yield that $A+B\Psi_k(Q, R)=A+B\Psi_k(\hat{Q},\hat{R})$, and then applying the Woodbury identity  to \eqref{fm3b} yields that\begin{small}
\begin{equation}\label{t4}
\begin{aligned}
A+B\Psi_k(Q, R)&=\left(I-B\left(R + B^\top P_{k+1} B\right)^{-1} B^\top P_{k+1}\right)A\\
&=(I+BR^{-1}B^\top P_{k+1})^{-1}A\\
&=(I+B\hat{R}^{-1}B^\top \hat{P}_{k+1})^{-1}A
\end{aligned}
\end{equation}
\end{small}and $A+B\Psi_k(Q, R)$ is invertible. Assumption $\ref{as2}$ ($A$ is invertible, $B$ is full column rank) and \eqref{t4} lead to 
\begin{align}
R^{-1}B^\top P_{k+1} &= \hat{R}^{-1}B^\top \hat{P}_{k+1} \label{ep3}
\end{align}
Then $\exists\alpha\ne 0$ such that $\alpha Q = \hat{Q}$, $\alpha R =  \hat{R}$. The detailed proof of this is presented for Theorem 1 in  \citep{b1} and outlined as below. Substituting \eqref{fm4b} and \eqref{tt7} into \eqref{ep3} and using the invertibility of $A + B\Psi_k$ gives $R^{-1}B^\top A^\top P_{k+1} = \hat{R}^{-1}B^\top A^\top \hat{P}_{k+1}$. Induction then yields $R^{-1}B^\top A^{\top,k} Q = \hat{R}^{-1}B^\top A^{\top,k} \hat{Q}$ for $k = 0, \dots, K-1$.
Therefore, the matrix triplets $(Q,A,BR^{-1})$ and $(\hat{Q},A,B\hat{R}^{-1})$ represent different realizations of a virtual system.
Under Assumption \ref{as3}, a similarity transformation $\mathcal{T}$ exists with $A = \mathcal{T}A\mathcal{T}^{-1}$, $B\hat{R}^{-1} = \mathcal{T}B R^{-1}$, and $\hat{Q} = Q\mathcal{T}^{-1}$. Since $A\mathcal{T} = \mathcal{T}A$, it follows that $A$ and $\mathcal{T}$ are simultaneously triangularizable, i.e., $A = \Pi\Lambda\Pi^{-1}$, $\mathcal{T} = \Pi\Sigma\Pi^{-1}$, with upper triangular $\Lambda$, $\Sigma$. The symmetry of $\hat{Q}$ and $B\hat{R}^{-1}B^\top$ leads to $\mathcal{Q} = \Pi\Sigma\Pi^{-1} \mathcal{Q} \Pi\Sigma^{-1}\Pi^{-1}$. Hence the existence of $\alpha$ stated above follows from the condition (ii) by showing that $\Pi\Sigma\Pi^{-1}$ or $\Sigma$ should be a scaled identity matrix.

With $\alpha R=\hat{R}$, it can be derived from \eqref{ep3} that
\begin{align}
\alpha B^\top P_{k+1} =  B^\top \hat{P}_{k+1},\quad \alpha  P_{k+1}B &= \hat{P}_{k+1}B \label{ep7}
\end{align}
Define $\Delta\eta_{k+1} = \alpha\eta_{k+1} - \hat{\eta}_{k+1}$, $\Delta q = \Delta\eta_{K+1} = \alpha q -\hat{q}$. Substituting \eqref{fm3c}, \eqref{ep7} into \eqref{t2} yields $B^\top\Delta\eta_{k+1} = \bm{0}$. On the other hand, equations \eqref{fm4a}, \eqref{t2}, \eqref{ep7} give $\Delta\eta_{k} = \Delta q + A^\top\Delta\eta_{k+1}$. Thus $\Delta q^\top \begin{bmatrix}B&AB&\cdots&A^KB\end{bmatrix}=\bm{0}$. Controllability and $K\ge 2n> n-p$ imply $\Delta q = \bm{0}$, so $\alpha q = \hat{q}$. As a result, it follows from $\alpha Q = \hat{Q}$, $q = -{Q}d$, and $\mathcal{P}^{\text{ker}({Q})}{d}=\bm{0}$ that $\hat{d} = -\hat{Q}^\dagger\hat{q}={Q}^\dagger {Q}{d}={d}$.  

Finally, the conclusion holds under \eqref{tt0f} as well, instead of the strict conditions that both $\hat{R}$ and $\hat{R}+B^\top \hat{P}_{k+1}B$ are either positive or negative definite. To see this, by Assumption \ref{as1} and Lemma \ref{lemma1}, $R\in\mathbb{S}_{++}^p$, ${R}+B^\top {P}_{k+1}B\in\mathbb{S}_{++}^p$, then the symmetry implies existence of $\epsilon\ne 0$ such that $R+\epsilon\hat{R}\in\mathbb{S}_{++}^p$, $R+\epsilon\hat{R}+B^\top (P_{k+1}+\epsilon\hat{P}_{k+1})B\in\mathbb{S}_{++}^p$ for any solution satisfying \eqref{tt0f} that $\hat{P}_{k}\in \mathbb{S}^n$ and $\hat{R}\in \mathbb{S}^p\setminus\{\bm{0}\}$. Since  $Q+\epsilon\hat{Q}$, $R+\epsilon\hat{R}$, $q+\epsilon\hat{q}$, $\hat{P}_k+\epsilon\hat{P}_k$, $\eta_k+\epsilon\hat{\eta}_k$ must solve \eqref{tt0}, with the conclusion above,  $\exists\alpha\ne 0$ satisfies ${R}+B^\top {P}_{k+1}B+\epsilon(\hat{R}+B^\top\hat{P}_{k+1}B)=\alpha({R}+B^\top {P}_{k+1}B)$, $R+\epsilon\hat{R}=\alpha{R}$. Since $\hat{R}\ne \bm{0}$ and $\epsilon\ne 0$, $\alpha\ne 1$. Let $\delta=(\alpha-1)/\epsilon\ne 0$. Then $\hat{R}+B^\top\hat{P}_{k+1}B=\delta({R}+B^\top {P}_{k+1}B)$, $\hat{R}=\delta{R}$, replacing the strict definiteness with \eqref{tt0f}. This completes the proof of the theorem.
\end{pf}
\begin{rem}
Condition (ii) is not a necessary condition, but simplifies the presentation and aligns with previous results \citep{b1}, including system-theoretic interpretations therein.
{It would be left for future work to develop conditions that can be validated prior to applying the algorithm to ensure that the identifiability assumptions are indeed satisfied.
}
\end{rem}

Theorem \ref{thm1}'s condition (iii) is a persistently exciting condition, verifiable and typically guaranteed.

\begin{lem}
For $M\ge n+1$ under Assumptions \ref{as1},\ref{as2} and \ref{as3}, $Y_{k}$ has full row rank if $X_k$ has full row rank.
\end{lem}
\begin{pf}
Equations (\ref{fm2}), (\ref{fm3a}) and (\ref{tt0}) yield:
$Y_{k}=\begin{bmatrix}A+B\Psi_k(Q,R)&B\psi_k(Q,R,d)\\\bm{0}&1\end{bmatrix}X_k$.
With invertibility of \( A + B\Psi_k(Q, R) \) shown in the proof of Theorem~\ref{thm1}, Schur complement conditions imply invertibility of the augmented closed-loop system matrix. Thus, full row rank of $X_k$ ensures that of $Y_{k}$. This completes the proof.
\end{pf}
\begin{rem}
When $X_k$ has full row rank,
$X_{k+1}=Y_k+\begin{bmatrix}w_{k,1}&w_{k,2}&\cdots&w_{k,M}\\0&0&\cdots&0\end{bmatrix}$
almost surely has full row rank if $w_{k,i}$ is uncorrelated over $k,i$ and with initial states $x_{0,i}$. Given $x_0\sim \mathcal{N}(0,\Sigma_0)$, $X_0$ almost surely has full row rank. Condition (iii) in Theorem \ref{thm1} thus holds almost surely by induction.
\end{rem}

\section{Identification algorithm}\label{sec4}
This section introduces a closed-form identification algorithm derived from equation (\ref{tt0}). Following this, the identification error is analyzed in the presence of observation noise, and the relevant assumptions are included.
\subsection{Reformulation of linear equations}
For simplicity, the data is first grouped. Suppose that the \( M \) trajectories are evenly divided into \( T \) groups, each containing \( s=\lfloor M/T\rfloor\ge n+1 \) trajectories, discarding any redundant data. 
For $r=1,2,\cdots,T$, define
\begin{small}
\begin{align}\label{gp1}
X_{k,r} = 
\begin{bmatrix}
x_{k,r_1}  & \cdots & x_{k,r_s} \\
1  & \cdots & 1 
\end{bmatrix},
\ U_{k,r} = 
\begin{bmatrix}
u_{k,r_1} & \cdots & u_{k,r_s}
\end{bmatrix} 
\end{align}  
\end{small}where indices of the data are non-repeating. 
Then, let 
\begin{align}\label{gp2}
X_k=\frac{1}{T}\sum_{r=1}^TX_{k,r},\ U_k=\frac{1}{T}\sum_{r=1}^TU_{k,r}
\end{align}
and perform a verification to ensure that \( X_{k} \) has full row rank, for all \( k = 0, \dots, K \).
Reviewing Theorem \ref{thm1}'s proof, replacing \(X_k\) and \(U_k\) with their redefined versions from (\ref{gp2}) preserves the conclusion. 
Ignoring observation noise, \(X_k\), \(U_k\), and \(Y_k\) satisfy (\ref{tt0}), vectorized as  
\begin{small}  
\begin{subequations}\label{eq30}  
\begin{align}  
&\Phi_{k}^B (Y_{k}\otimes I_n)^\top \begin{bmatrix}  
    vec(P_{k+1})\\\eta_{k+1}  
\end{bmatrix}+(U_{k}\otimes I_p)^\top vec(R)=\bm{0}\\  
&\Phi_{k}^A (Y_{k}\otimes I_n)^\top \begin{bmatrix}  
    vec(P_{k+1})\\\eta_{k+1}  
\end{bmatrix}+(X_{k}\otimes I_n)^\top \begin{bmatrix}  
    vec(Q-P_{k})\\q-\eta_{k}  
\end{bmatrix}=\bm{0}  
\end{align}  
\end{subequations}  
\end{small}where hats $\hat{\cdot}$ are omitted as the ground truth solves (5), with $\Phi_{k}^B=I_s\otimes B^\top$ and $\Phi_{k}^A=I_s\otimes A^\top$. 
Given $Q,R,P_k\in\mathbb{S}^n$, (\ref{eq30}) can be simplified by half-vectorization via  
\begin{align}  
\begin{bmatrix}  
    vec(P_{k})\\\eta_{k}  
\end{bmatrix}=\bar{\mathcal{D}}_n\begin{bmatrix}  
    vech(P_{k})\\\eta_{k}  
\end{bmatrix},\quad\bar{\mathcal{D}}_n=\begin{bmatrix}\mathcal{D}_n&\\&I_n\end{bmatrix}\label{r1d}  
\end{align}  
for $k=1,2,\cdots,K+1$, and $vec(R)=\mathcal{D}_p vech(R)$, with duplication matrices $\mathcal{D}_n,\mathcal{D}_p$ possessing full column rank. To determine $(Q, R, q)$ satisfying (\ref{tt0}), define
\begin{align*}
        \theta=\begin{bmatrix}(vech(R))^\top&(vech(Q))^\top&q^\top\end{bmatrix}^\top,\ \gamma_k=\begin{bmatrix}vech(P_k)\\\eta_k\end{bmatrix}
\end{align*}
Then, a compact form of \eqref{eq30} is: $\begin{bmatrix}\Phi_{k}^u&\bm{0}\end{bmatrix}\theta+\Phi_{k}^{B,y}\gamma_{k+1}=\bm{0}$, $\begin{bmatrix}\bm{0}&\Phi_{k}^x\end{bmatrix}\theta+\Phi_{k}^{A,y}\gamma_{k+1}=\Phi_{k}^x\gamma_k$ and $\begin{bmatrix}\bm{0}&I_{n(n+3)/2}\end{bmatrix}\theta=\gamma_{K+1}$, 
where $\Phi_{k}^{B,y}=\Phi_{k}^B\Phi_{k}^y$, $\Phi_{k}^{A,y}=\Phi_{k}^A\Phi_{k}^y$ and\begin{small}
\begin{equation}\label{r1}  
\Phi_{k}^u=(U_{k}\otimes I_p)^\top \mathcal{D}_p, \Phi_{k}^y=(Y_{k}\otimes I_n)^\top \bar{\mathcal{D}}_n,
\Phi_{k}^x=\left(X_{k}\otimes I_n\right)^\top\bar{\mathcal{D}}_n
\end{equation}    
\end{small}To estimate $\theta$, eliminate $\gamma_k$ with $\gamma=\begin{bmatrix}\gamma_1^\top&\gamma_2^\top&\cdots&\gamma_{{K}}^\top\end{bmatrix}^\top$, yielding  
\begin{align}\label{r2}  
    \underbrace{\begin{bmatrix}\Phi^U&\Phi^{Y_0}&\Phi^{Y_1}\\\bm{0}&\Phi^{X}&\Phi^{Y_2}\end{bmatrix}}_{:=S}  
    \underbrace{\begin{bmatrix}\theta\\\gamma\end{bmatrix}}_{:=z}=\bm{0}  
\end{align}  
with the components of $S$ are given by\begin{small}  
\begin{equation}  
\label{r4}  
\begin{aligned}  
&\Phi^U=\begin{bmatrix}  
\Phi_{0}^u\\  
\vdots\\  
\Phi_{K}^u  
\end{bmatrix},  
\Phi^{Y_0}=\begin{bmatrix}\bm{0}\\\vdots\\\bm{0}\\\Phi_K^{B,y}\end{bmatrix},  
\Phi^{Y_1}=  
\begin{bmatrix}  
\Phi_0^{B,y}&&	\\  
  &\ddots &\\  
&& \Phi_{K-1}^{B,y}\\  
\bm{0}&\cdots&\bm{0}		  
\end{bmatrix},  
\\  
&
\Phi^{X}=\begin{bmatrix}  
\Phi_{1}^x\\  
\vdots\\  
\Phi_{K-1}^{x}	\\  
\Phi_{K}^x+\Phi_K^{A,y}  
\end{bmatrix}, 
\Phi^{Y_2}=  
\begin{bmatrix}  
-\Phi_{1}^x&\Phi_{1}^{A,y}&&\\  
&\ddots&\ddots&\\  
&			   	&-\Phi_{K-1}^x	  	&\Phi_{K-1}^{A,y}\\    
			   	&&	  	 	&-\Phi_{K}^x	  
\end{bmatrix}  
\end{aligned}  
\end{equation}      
\end{small}where blanks denote zero components. Since $\mathcal{D}_n$ has full column rank, $\bar{\mathcal{D}}_n$ does by (\ref{r1d}), so $\Phi_{k}^x$ has full column rank per (\ref{r1}). By (\ref{r4}), $\Phi^{Y_2}$ has full column rank, giving $\gamma=-(\Phi^{Y_2})^\dagger\Phi^{X}\gamma_{K+1}$ if and only if $\Phi^{X}\gamma_{K+1}+\Phi^{Y_2}\gamma=\bm{0}$. Substituting $\gamma$ in terms of $\theta$, (\ref{r2}) becomes  
\begin{equation}\label{r3}
        Z\theta=\bm{0},\quad
        Z=\begin{bmatrix}\Phi^U&\Phi^{Y_0}-\Phi^{Y_1}(\Phi^{Y_2})^\dagger\Phi^{X}\\\bm{0}&\left(I_{snK}-\Phi^{Y_2}(\Phi^{Y_2})^\dagger\right)\Phi^{X}\end{bmatrix}
\end{equation}  
Solution $\theta$ is obtained from (\ref{r3}); homogeneity implies that its solution set is determined by the kernel of $Z$.

Theorem~\ref{thm1} shows that solving \eqref{tt0} yields an estimate on the half-line from the origin to the true parameters, with signs adjusted per Assumption~\ref{as1}.  
Without the constraint \(R \neq \bm{0}\) in \eqref{tt0f}, the solution lies on the full line through the origin, as one directly solves \(Z\theta = \bm{0}\).  
Since \(Z\) is a tall matrix whose column rank is deficient by one, the right singular vector corresponding to its unique zero singular value—once normalized—gives such an estimate. Therefore, reformulating the optimality conditions leads to the following theoretical statement.

\begin{prop}
    Under the conditions of Theorem~\ref{thm1}, solving \eqref{r3} with \(\theta \neq \bm{0}\) yields the same solutions as \eqref{tt0}.
\end{prop}

For large-scale linear equations, singular value decomposition (SVD) is computationally inefficient. To enhance efficiency and ensure normalization, an additional constraint \( C\theta = c \) with \( c \neq \bm{0} \) is imposed, yielding the inhomogeneous linear equation 
\begin{equation}\label{r5p}
    H\theta = h,\quad
    H = \begin{bmatrix} C^\top & Z^\top \end{bmatrix}^\top,\quad
    h = \begin{bmatrix} c^\top & \bm{0} \end{bmatrix}^\top
\end{equation}
This provides a canonical solution; e.g., setting \(C = [1\ 0\ \cdots\ 0]\) and \(c = 1\) normalizes the estimate. Alternatively, \(C\) and \(c\) may encode prior information.

Since \(Z\) is rank-deficient by one, adding more than one independent constraint generally renders the system infeasible unless consistent with \(Z\theta = \bm{0}\).  Therefore, the following assumption is needed for further analysis.

\begin{assum}\label{asC}
    The constraint \(C\theta = c\) in \eqref{r5p}, where \(c\neq \bm{0}\),  is consistent with \(Z\theta = \bm{0}\).
\end{assum}

This is reasonable when structural information on \(\theta\) is available (e.g., \eqref{m7} and \eqref{m6} introduced later). Under Assumption~\ref{asC}, at least one row of \(C\) is linearly independent of the rows of \(Z\), so \(H\) has full column rank and \(\theta = H^\dagger h\) is uniquely determined.

The implementation process of the algorithm is summarized in Algorithm \ref{alg1}.
\begin{small}
    \begin{algorithm}[t!]
	\caption{Inverse LQT with unknown target state}
	\label{alg1}
	\begin{algorithmic}[0]
		\State{Input: $A,B,u_k$ and $x_k$.}
        \State {Output: General solution, \\\qquad\qquad i.e.  $(\alpha {Q},\alpha {R}, {d}+\mathcal{P}^{\text{ker}({Q})}\lambda),\forall\alpha>0,\lambda\in\mathbb{R}^n$.}	
	\end{algorithmic}
	\hrule
	\begin{algorithmic}[0]
	\State \textbf{Step 1:} Linear equations construction
	\State \textbf{(1)} Select  $T$ groups of the optimal control input and the corresponding state, and compute $(X_k,U_k,Y_k)$, such that $X_k$ has full row rank for each time  $k={0}, 1,\cdots,{K}$, with (\ref{gp1}) and (\ref{gp2});
	\State \textbf{(2)}  Calculate the intermediate variables $\Phi_{k}^u,\Phi_{k}^x,\Phi_{k}^y$ according to equation (\ref{r1});
	\State \textbf{(3)}  Construct the data matrices $\Phi^U$, $\Phi^X$, $\Phi^{Y_0}$, $\Phi^{Y_1}$ and $\Phi^{Y_2}$ according to equation (\ref{r4});
	\State \textbf{(4)}  Obtain the coefficient matrix $Z$ according to (\ref{r3}).
	\end{algorithmic}
	\hrule
	\begin{algorithmic}[0]
	\State \textbf{Step 2:} Particular solution acquisition
	{\State \textbf{(1)}  Compute $\hat{\theta}$ from (\ref{r5p});
	\State \textbf{(2)}  Restructure $(\hat{Q}, \hat{R}, \hat{q})$ with $\hat{\theta}$, and then $\hat{d}=-\hat{Q}^\dagger\hat{q}$.}		
	\end{algorithmic}
\end{algorithm}
\end{small}
\subsection{Error analysis}\label{sec4b}
The influence of observation noise is examined to evaluate the statistical properties of the solution. Variables introduced earlier are now considered as true values, while their observed noisy counterparts are represented as \( \tilde{\cdot} \), with observation noise denoted by \( \Delta{\cdot} \). 
Unless otherwise stated, the following assumption holds.  
\begin{assum}\label{as4}  
Observed \(\tilde{x}_{k,i}\) and \(\tilde{u}_{k,i}\) include zero-mean additive noise \(\Delta x_{k,i}\) and \(\Delta u_{k,i}\), uncorrelated across \(k\) and \(i\), with bounded second moments:  
\(\tilde{x}_{k,i}=x_{k,i}+\Delta x_{k,i}\), \(\tilde{u}_{k,i}=u_{k,i}+\Delta u_{k,i}\),  
where \(\mathbb{E}\{\Delta x_{k,i}\}=\bm{0}\), \(\|\mathbb{E}\{\Delta x_{k,i}\Delta x_{k,i}^\top\}\|<\infty\), \(\mathbb{E}\{\Delta u_{k,i}\}=\bm{0}\), \(\|\mathbb{E}\{\Delta u_{k,i}\Delta u_{k,i}^\top\}\|<\infty\), and \(\mathbb{E}\{\Delta x_{k,i}\Delta x_{k',i'}^\top\}=\bm{0}\), \(\mathbb{E}\{\Delta u_{k,i}\Delta u_{k',i'}^\top\}=\bm{0}\) for \(i\ne i'\) or \(k\ne k'\).  
\end{assum}  

To utilize the assumption more explicitly, the analysis is based on (\ref{r2}). Under Theorem 1's conditions, the IOC problem can be solved by the following equation.  
\begin{align}\label{eaa1}  
\Omega z=h,\quad\Omega=\begin{bmatrix}  
            \begin{bmatrix}C&\bm{0}\end{bmatrix}^\top&S^\top  
        \end{bmatrix}^\top  
\end{align}  
with $z$ defined in \eqref{r2} and \(h\) defined in \eqref{r5p}. Analogous to \(H\), \(\Omega\) has full column rank for \(c\ne\bm{0}\). The noisy matrix is  
\begin{align}\label{ea6}  
\tilde{\Omega}=\Omega+\Delta{\Omega}  
=\Omega  
+\begin{bmatrix}  
            \begin{bmatrix}\bm{0}&\bm{0}\end{bmatrix}^\top&\Delta S^\top  
        \end{bmatrix}^\top  
\end{align}  
Per (\ref{r1})-(\ref{r4}), \(\Delta S\) and \(\Delta\Omega\) are linear in \(\frac{1}{T}\sum_{r=1}^T \Delta X_{k,r}\) and \(\frac{1}{T}\sum_{r=1}^T \Delta U_{k,r}\). Hence, \(\Delta \Omega=\frac{1}{T}\sum_{r=1}^T \Delta \Omega_{r}\), where \(\Delta \Omega_r\) are linear in \(\Delta X_{k,r}\) and \(\Delta U_{k,r}\). By Assumption \ref{as4},  
\(\mathbb{E}\{\Delta \Omega_r\} = \bm{0},\mathbb{E}\{\|\Delta\Omega_r\|^2\}<\infty,  
\mathbb{E}\{\Delta\Omega_r^\top\Delta\Omega_{r'}\}=\bm{0}\) for \(r\ne r'\), yielding  
\begin{equation}\label{eq:edo}  
    \begin{split}  
    \mathbb{E}\{\|\Delta\Omega\|\}\le\sqrt{\mathbb{E}\{\|\Delta\Omega\|^2\}}=\sqrt{\mathbb{E}\{\|\frac{1}{T}\sum_{r=1}^T \Delta \Omega_{r}\|^2\}}\\  
    =\frac{1}{T}\sqrt{\sum_{r=1}^T \mathbb{E}\{\|\Delta \Omega_{r}\|^2\}}=\mathcal{O}(T^{-0.5})  
\end{split}  
\end{equation}   
as \(T\to\infty\). For clarity, the following definition is proposed, resembling prior IOC work (\citep{b1,bz5}) but differing to make SNR finite.  
\begin{align}\label{sim2}  
\text{SNR}=10\text{log}\frac{\left\|\Omega\right\|^2}{T\left\|\Delta{\Omega}\right\|^2}  
\end{align}  
with SNR \(=10\text{log}(\|\Omega\|^2/(T\mathcal{O}(T^{-1})))=\mathcal{O}(1)\) as \(T\to\infty\), due to \eqref{eq:edo}. {Under this definition, the SNR, though dependent on $T$, could be treated as constant for a given dataset. The analysis then focuses on how the data horizon $T$ affects the upper bound of the relative error.}
Define the estimation errors of $\tilde{\theta}$ and $\tilde{z}$ as 
\begin{align}\label{ea4}
\Delta{\theta}=\tilde{\theta}-{\theta},\quad\Delta{z}=\tilde{z}-{z}
\end{align}
where $\tilde{\theta}$, $\tilde{\gamma}$ and $\tilde{z}$ are the solutions to 
\begin{align}
    \tilde{\Omega}\tilde{z}=h\label{ea5},
\quad\tilde{z}=\begin{bmatrix}{\tilde{\theta}}^\top&{\tilde{\gamma}}^\top\end{bmatrix}^\top
\end{align}
and ${\theta}$ and $z$ are the ground truth. 
The above definitions allows us to analyze the statistical properties of the solutions to the concerned IOC problem. {Specifically, although equation (\ref{ea5}) does not hold exactly under observation noise, it is solvable in the least-squares sense, with the resulting residual assumed negligible.  In the following theorem, the asymptotic behavior is analyzed and an upper bound on the estimation error as $T \to \infty$ is provided.}

\begin{thm}
\label{thm2}
Given $T$ groups of observed optimal control trajectories $\left\{\tilde{X}_{k,r}, \tilde{U}_{k,r}\right\}_{k=0,r=1}^{K,T}$ defined in (\ref{gp1}), each containing $s \geq n+1$ noisy state-input pairs $\left\{\tilde{x}_{k,i},\tilde{u}_{k,i}\right\}_{k=0,i=r_1}^{K,r_s}$, assume: (i) Assumptions \ref{as1}, \ref{as2}, \ref{as3}, \ref{asC} and \ref{as4} hold; (ii) $\mathcal{Q}A \neq A\mathcal{Q}$; (iii) Each $X_k$ in (\ref{gp2}) has full row rank; (iv) Equation \eqref{ea5} can be solved exactly. Then $\tilde{\theta}$ consistently estimates $\theta$, and as $T\to\infty$:
\begin{align}\label{sim1}
\frac{\|\Delta\theta\|}{\|\theta\|}\leq
T^{-0.5}\text{cond}(\Omega)10^{-0.05\text{SNR}}\frac{\|z\|}{\|\theta\|}+\mathcal{O}\left(T^{-1}\right)
\end{align}
where $\text{cond}(\cdot)$ denotes the $L_2$-norm condition number.
\end{thm}

\begin{pf}
By singular value perturbation theory (c.f. Corollary 7.3.5 in \citep{horn2012matrix}), \eqref{eq:edo} implies that when $T\to\infty$,
\begin{align}\label{eq:svp}
|\sigma_{\min}(\tilde{\Omega}) - \sigma_{\min}(\Omega)| \leq \|\Delta\Omega\| = \mathcal{O}(T^{-0.5})
\end{align}
Since $\sigma_{\min}(\Omega)>0$, $\sigma_{\min}(\tilde{\Omega})$ is also bounded away from zero for sufficiently large $T$. Then  (\ref{ea5}) gives the boundedness of $\tilde{z}$.
Besides, it follows from (\ref{eaa1}), (\ref{ea6}),   (\ref{ea4}) and (\ref{ea5}) that
\begin{align}\label{bs4}
\Omega z = h = \tilde{\Omega}\tilde{z} = (\Omega + \Delta\Omega)(z + \Delta z)
\end{align}
since both noiseless and noisy data share the identical inhomogeneous term $h$. For zero-mean observation noise,
\begin{align}
\lim_{T\to\infty}\Delta{\Omega}=\lim_{T\to\infty}\frac{1}{T}\sum_{r=1}^T \Delta \Omega_{r}=\mathbb{E}\{\Delta \Omega_r\}= \bm{0}
\end{align}
 with which the equation \eqref{bs4} yields the following result
\begin{align}
\lim_{T\to\infty}\Omega\Delta z=\lim_{T\to\infty}(\Omega z-\Omega z-\Delta\Omega(z+\Delta z))=\bm{0}
\end{align}
with bounded $z+\Delta z=\tilde{z}$. Since $\Omega$ has full column rank,
\begin{align}
\lim_{T\to\infty}\Delta z = \bm{0}, \quad \mathbb{E}\{\tilde{z}\}=\lim_{T\to\infty}z+\Delta{z}=z
\end{align}
confirming $\tilde{\theta}$'s consistency. From (\ref{bs4}), $\Delta z = -\tilde{\Omega}^\dagger\Delta\Omega z$, and with norm compatibility, $\|\Delta z\| \leq \|\tilde{\Omega}^\dagger\|\|\Delta\Omega\|\|z\|$, or
\begin{align}\label{ea8}
\|\Delta z\|/\|z\|
\leq \|\tilde{\Omega}^\dagger\|\|\Delta\Omega\| = \mathcal{O}(T^{-0.5})
\end{align}
as $\|\Delta\Omega\|=\mathcal{O}(T^{-0.5})$ and $\|\tilde{\Omega}^\dagger\|=\mathcal{O}(1)$ when $T\to\infty$, for the monotonicity property of limits. 

Furthermore, the spectral norms of the pseudo-inverses are given by $\|\Omega^\dagger\| = {1}/{\sigma_{\min}(\Omega)}, \quad \|\tilde{\Omega}^\dagger\| = {1}/{\sigma_{\min}(\tilde{\Omega})}$.
Since both $\sigma_{\min}(\Omega)$ and $\sigma_{\min}(\tilde{\Omega})$ are bounded away from zero for sufficiently large $T$, \eqref{eq:svp} leads to
\begin{align}\label{eq:bondo}
\left| \|\tilde{\Omega}^\dagger\| - \|\Omega^\dagger\| \right| = \frac{\left| \sigma_{\min}(\tilde{\Omega}) - \sigma_{\min}(\Omega) \right|}{\sigma_{\min}(\tilde{\Omega}) \sigma_{\min}(\Omega)}\le \mathcal{O}\left(T^{-0.5}\right)
\end{align}
By (\ref{eq:edo}), (\ref{sim2}), (\ref{ea8}), (\ref{eq:bondo}), and $\text{cond}(\Omega)=\|\Omega^\dagger\|\|\Omega\|$,
\begin{equation}
    \begin{aligned}\label{ea9}
\frac{\|\Delta{z}\|}{\|z\|}&\leq \|\tilde{\Omega}^\dagger\|\|\Delta\Omega\|\\&\le
\left(\|\Omega^\dagger\| + \mathcal{O}\left(T^{-0.5}\right)\right)\left\|\Delta{\Omega}\right\|\\
&=\|{\Omega}^\dagger\|\left\|\Omega\right\|\frac{\left\|\Delta{\Omega}\right\|}{\left\|\Omega\right\|}+\mathcal{O}\left(T^{-0.5}\left\|\Delta{\Omega}\right\|\right)\\
 &= T^{-0.5}\text{cond}(\Omega)10^{-0.05\text{SNR}}+\mathcal{O}\left(T^{-1}\right)
\end{aligned}
\end{equation}
According to (\ref{r2}) and (\ref{ea4}), it is clear that
$\frac{\|\Delta\theta\|}{\|\theta\|} = \frac{\|z\|}{\|\theta\|}\frac{\|\Delta\theta\|}{\|z\|} \leq \frac{\|z\|}{\|\theta\|}\frac{\|\Delta z\|}{\|z\|}$.
Thus (\ref{sim1}) holds when $T\to\infty$. This completes the proof.\end{pf}
\begin{rem}
As \( s < \infty \), the convergence rate is \( \mathcal{O}(T^{-0.5}) = \mathcal{O}(M^{-0.5}) \), with \( M \) the total observed trajectories. Additionally, \( \text{cond}(\Omega) \geq 1 \) reflects grouped true data sensitivity to noise, \( 10^{-0.05\text{SNR}} \) represents noise magnitude, and \( \|z\|/\|\theta\| \) indicates parameter sensitivity.
\end{rem}

For small \( T \), the bound resembles \eqref{sim1}, but it is less direct than the classical results: \( \frac{\|\Delta \theta\|}{\|\theta\|} \leq \frac{\varepsilon}{1-\varepsilon} \frac{\|z\|}{\|\theta\|} \) where \( \varepsilon = \text{cond}(\Omega) \frac{\|\Delta \Omega\|}{\|\Omega\|} \), circumventing computation of \( \|\tilde{\Omega}^\dagger\| \). Indeed, estimating \( \text{cond}(\Omega) \) accurately for small \( T \) is challenging, complicating its use as a bound.

\section{Multi-agent LQT formation  under fixed but unknown topology}\label{sec5}
This section specializes the proposed framework for multi-agent LQT under fixed but unknown topology, exploring networked agents' coordination patterns and intents. Consider $N$ agents. Each agent $j$ has state $x_{k}^j\in\mathbb{R}^{n_0}$ and input $u_{k}^j\in\mathbb{R}^{p_j}$, forming the cluster state $x_k=\begin{bmatrix}x_{k}^{1,\top}&\cdots&x_{k}^{N,\top}\end{bmatrix}\in\mathbb{R}^n$ and input $u_k=\begin{bmatrix}u_{k}^{1,\top}&\cdots&u_{k}^{N,\top}\end{bmatrix}\in\mathbb{R}^p$,
where $n={Nn_0}$, $p={\sum_{j=1}^N{p_j}}$. The cluster dynamics are $x_{k+1}^j=A_{j}x_{k}^j+B_ju_{k}^j+w_{k}^j$
with intents $d^j\in\mathbb{R}^{n_0}$ aggregated into $d$. Multi-agent behavior follows (\ref{fm1}), with $A=\text{diag}(\{A_j\}_{j=1}^N)$, $B=\text{diag}(\{B_j\}_{j=1}^N)$ and $d=\begin{bmatrix}d^{1,\top}&\cdots&d^{N,\top}\end{bmatrix}^\top$.
The LQT policy under unknown topology follows (\ref{fm2}) by minimizing (\ref{fm1a}), where
\begin{align}\label{m1}
Q={(L+I_N)}\otimes I_{n_0},\ R=\text{diag}(\phi),\ \phi\in\mathbb{R}^p
\end{align}
Here, $L$ is the Laplacian for a connected undirected graph \citep{bz3}, satisfying:
(i) $L\in\mathbb{S}_+^N$; (ii) $\mathbf{1}_N^\top L=\bm{0}$.

\begin{prob}\label{prob2}
Consider the multi-agent LQT formation policy $\pi_k$ in (\ref{fm2}) under fixed  but unknown topology, with $Q$ and $R$ being defined in (\ref{m1}).  Given $M$ optimal state and input trajectories with length $K$, i.e., $\{x_{k,i},u_{k,i}\}_{k=\bm{0},i=1}^{K,M}$,
the concerned IOC problem aims to find $(L,\phi,d)$ satisfying:
$u_{k,i}=\pi_k\left({(L+I_N)}\otimes I_{n_0},\text{diag}(\phi),d,x_{k,i}\right)$.
\end{prob}

Structural priors ensure unique identifiability of $(L, \phi, d)$. As in Theorem \ref{thm1} and Algorithm \ref{alg1}, unique $(Q, R)$ requires constraints $C\theta=c$ ($c \neq \bm{0}$), while unique $d$ requires $\mathcal{P}^{\text{ker}(Q)}d=\bm{0}$.

\begin{lem}\citep{bz4}.\label{lemma3}
For $L+I_N$, there exist $\mathcal{F},\mathcal{H}$ such that:
\begin{align}\label{m4}
\text{vec}({(L+I_N)}\otimes I_{n_0})=(I_N\otimes \mathcal{F})\text{vec}({(L+I_N)})
\end{align}
where $\mathcal{F}=(\mathcal{H}\otimes I_{n_0})(I_N\otimes \text{vec}(I_{n_0}))$, and $I_N\otimes \mathcal{F}$ has full column rank. $\mathcal{H}$ is the commutation matrix.
\end{lem}

\begin{cor}\label{coro2}
For Problem \ref{prob2} with $L\in\mathbb{S}_+^n$ and $\mathbf{1}_N^\top L=\bm{0}$, define $C$ as:
\begin{align}\label{m7}
C=\begin{bmatrix}0_{N\times p(p+1)/2}&(I_N\otimes\mathbf{1}_N^\top)(I_N\otimes \mathcal{F})^\dagger\mathcal{D}_n&0_{N\times n}\end{bmatrix}
\end{align}
with $c=\mathbf{1}_N$. Under Theorem 1's conditions, solution $(\tilde{L}, \tilde{\phi}, \tilde{d})$ to (\ref{r5p}) satisfies $\tilde{L}= L$, $\tilde{\phi}= \phi$, $\tilde{d}=d$.
\end{cor}
\begin{pf}
Show $\tilde{Q}=Q$, $\tilde{R}=R$ via $C\theta=c$:
$C\theta=(I_N\otimes\mathbf{1}_N^\top)(I_N\otimes \mathcal{F})^\dagger \text{vec}({(L+I_N)}\otimes I_{n_0})=(I_N\otimes\mathbf{1}_N^\top)\text{vec}({L+I_N})=\text{vec}(\mathbf{1}_N^\top (L+I_N))=\mathbf{1}_N=c.$
Thus $\tilde{L}=L$, $\tilde{\phi}=\phi$ by Theorem \ref{thm1} and (\ref{m1}). Since $L\in\mathbb{S}_+^N$ implies $Q=(L+I)\otimes I_{n_0}\in\mathbb{S}_{++}^n$, $\mathcal{P}^{\text{ker}(Q)}d=\bm{0}$ gives $\tilde{d}=d$.
\end{pf}
Let 
$
\bar{\theta}:=\begin{bmatrix}\phi^\top&(\text{vech}(L+I_N))^\top&q^\top\end{bmatrix}^\top
$
and according to (\ref{m1}) and (\ref{m4}), it can be obtained that
{\begin{align}\label{m5}
\theta=\begin{bmatrix}vech(\text{diag}(\phi))\\\mathcal{D}_n^\dagger(I_N\otimes \mathcal{F})\mathcal{D}_Nvech(L+I_N)\\q\end{bmatrix}
\end{align}}then $\theta$ can be reconstructed after the identification of $\bar{\theta}$. 

Redefine
\begin{subequations}\label{m6}
\begin{align}
U_{k,r} &= 
\begin{bmatrix}
\text{diag}(u_{k,r_1}) &\text{diag}( u_{k,r_2}) & \cdots &\text{diag} (u_{k,r_s})
\end{bmatrix} \\
\bar{\mathcal{D}}_n&=\begin{bmatrix}(I_N\otimes \mathcal{F})\mathcal{D}_N&\\&I_n\end{bmatrix}\\
C&=\begin{bmatrix}0_{N\times p}&(I_N\otimes\mathbf{1}_N^\top)\mathcal{D}_N&0_{N\times n}\end{bmatrix}
\end{align}
\end{subequations}
with $c=\mathbf{1}_N$. Then $\bar{\theta}$ is estimated via Algorithm \ref{alg1}, and $\hat{\theta}$ is reconstructed by (\ref{m5}). Estimating $\hat{\theta}$ is more efficient than acquiring $\tilde{\theta}$ per Corollary \ref{coro2}, due to reduced dimensionality of the solution space. Additionally, the approach typically improves accuracy by precisely fulfilling structural prior knowledge under observation noise.
\begin{prop}\label{thm4}
Suppose that the observation noise is zero-mean with finite second moment. Let $\hat{\theta}$ denote the estimate by Algorithm \ref{alg1} with (\ref{m5})-(\ref{m6}), and $\tilde{\theta}$ be estimated from (\ref{gp1}), (\ref{r1d}) and (\ref{m7}), then 
\begin{align}
\|\hat{\theta}-{\theta}\|^2=\|\tilde{\theta}-\theta\|^2-\|\tilde{\theta}-\hat{\theta}\|^2
\end{align}
where $\theta$ denotes the true value.
\end{prop}
\begin{pf}
Following (\ref{m5})-(\ref{m6}), the procedure of obtaining $\hat{\theta}$ ensures that the duplicate and zero elements in $Q = {(L+I_N)} \otimes I_{n_0},R=\text{diag}(\phi)$ adhere to their structural characteristics, which can be expressed using the linear constraint $V\theta=v$. The resulting  $\hat{\theta}$ strictly satisfies this constraint, i.e. $V\hat{\theta}=v$.
Denote $\Delta Z$ as the noise term w.r.t $Z$ defined in (\ref{r3}), then $\hat{\theta}$ is the solution to the following optimization problem with $C$ defined in (\ref{m7}).\begin{small}
\begin{equation}
\begin{array}{c@{\quad}c}
    \min_{\hat{\theta}} \|\tilde{H}\hat{\theta}-h\|^2 & 
    \text{s.t.} \left\{
    \begin{array}{l}
        V\hat{\theta} = V\theta = v \\
        \tilde{H} = \begin{bmatrix} C^\top & {Z}^\top + \Delta Z^\top \end{bmatrix}^\top \\
        h = \begin{bmatrix} \mathbf{1}_N^\top & \bm{0} \end{bmatrix}^\top
    \end{array}
    \right. \\
\end{array}
\end{equation}
\end{small}Since $\tilde{H}\tilde{\theta}=h$ and $\tilde{H}$ has full column rank, $\hat{\theta}$ is also the solution of the following optimization problem
\begin{equation}\label{pe8}
\begin{array}{c@{\quad}c}
    \min_{\hat{\theta}} \|\hat{\theta}-\tilde{\theta}\|^2 & 
    \text{s.t.}\ V\hat{\theta}=V\theta
\end{array}
\end{equation}
in which strong duality holds and $(\hat{\theta},\mu)$ are a primal and dual optimal solution pair if and only if $\hat{\theta}$ is feasible and
$2(\hat{\theta}-\tilde{\theta})+V^\top\mu=\bm{0}$.
Given $(\hat{\theta}-\theta)^\top V^\top=\bm{0}$, it follows that
$2(\hat{\theta}-\theta)^\top(\tilde{\theta}-\hat{\theta})=\bm{0}$.
Therefore, it can be derived that
$\|\tilde{\theta}-\theta\|^2=\|\tilde{\theta}-\hat{\theta}+\hat{\theta}-{\theta}\|^2=\|\tilde{\theta}-\hat{\theta}\|^2+\|\hat{\theta}-{\theta}\|^2$.
This completes the proof of the proposition.
\end{pf}
\begin{rem}
Proposition \ref{thm4} indicates that $\|\hat \theta-\theta\|\leq \|\tilde\theta-\theta\|$, i.e., incorporating the topological prior knowledge yields  {an either equal or} more accurate estimate.
\end{rem}

The essence of this conclusion is rooted in the properties of projection.  From Theorem \ref{thm2} and Proposition \ref{thm4}, the relative error is $\mathcal{O}(T^{-0.5})$, and decreases with increasing $\|\tilde{\theta}-\hat{\theta}\|$, where $\hat{\theta}$ is estimated by Algorithm \ref{alg1} using (\ref{m5})-(\ref{m6}) under  conditions in Theorem \ref{thm2}. Using (\ref{m6}) instead of (\ref{gp1}), (\ref{r1d}), and (\ref{m7}), Algorithm \ref{alg1} applies to inverse identification of multi-agent LQT with fixed unknown topology, exploiting prior knowledge for solutions approximating the true value despite observation noise. In the multi-agent example, \eqref{m1} ensures \( P^{\text{ker}(Q)}d = \bm{0} \), guaranteeing unique identifiability of \( d \). For general problems, this condition may not be satisfied, thereby preventing the structural identifiability of $d$ from being guaranteed. Nevertheless, the proposed method yields a target state that characterizes system behavior through identifiable weighted components. In particular, if $d$ lies outside the range of $Q$, the unidentifiable components have no influence on agent behavior, allowing the cost function to be represented in a lower-dimensional form.
\section{Simulations}\label{sec6}
Two simulation examples validate the proposed IOC method: a vehicle-on-a-lever system and multiagent formation control. Without observation noise, comparisons are made with a baseline SDP method based on \citep{bz5}, demonstrating that the proposed method yields more accurate solutions efficiently for identical datasets. It is also compared with a computationally efficient baseline builds upon Pontryagin's maximum principle. With observation noise, the proposed algorithm is compared against the theoretical upper bound in Theorem \ref{thm2}, validating its consistency in identifying the weight matrices and the target state, even with unknown observation noise covariance.
\subsection{Baseline methods }

Two baseline methods are developed, which resemble the state-of-the-art approaches in \citep{bz5} and \citep{d3}. While \citep{bz5} assumes known process noise covariance but unknown control input—differing from this work—its modified version estimates parameters $(Q,R,q)$ via the following optimization problem.
\begin{small}
    \begin{align}
\begin{split}
	\min_{\substack{Q,P_k\in \mathbb{S}_+^n,\\R\in\mathbb{S}_{++}^p, \eta_{k},\xi_{k}}} 
	&\frac{1}{M}\sum_{k=0}^{{K}}\sum_{{i}=1}^{M}f_{k,i}(Q,R,d,P_k,\eta_k,P_{k+1},\eta_{k+1})
	 \label{op1}\\ 
	\text{s.t.} \qquad &\eta_{K}=q,P_{K}=Q,C\theta=c\\
	&\begin{bmatrix}
	\mathfrak{R}_k&\mathfrak{G}_k&\mathfrak{g}_{k}\\
	\mathfrak{G}_k^\top&Q+A^\top P_{k+1}A-P_k&\beta_{k}\\
	\mathfrak{g}_{k}^{\top}&\beta_{k}^{\top}&\xi_{k}\\
	\end{bmatrix}\in\mathbb{S}_{+}^{p+n+1}
\end{split}
\end{align}
\end{small}where $\mathfrak{R}_k=B^\top P_{k+1}B+R$, 
$\mathfrak{G}_k=B^\top P_{k+1}A$,
$\mathfrak{g}_k=B^\top\eta_{k+1}$,
$\beta_{k}=q+A^\top\eta_{k+1}-\eta_{k}$,
$y_{k,i}=Ax_{k,i}+Bu_{k,i}$ and $f_{k,i}(Q,R,d,P_k,\eta_k,P_{k+1},\eta_{k+1})=(Qx_{k,i}+2q)^\top x_{k,i}+(Ru_{k,i})^\top u_{k,i}+\xi_k+(P_{k+1}y_{k,i}+2\eta_{k+1})^\top y_{k,i}-(P_{k}x_{k,i}+2\eta_{k})^\top x_{k,i}$. In the optimization problem, $f_{k,i}$ is the residual from dynamic programming. Techniques in \citep{bz5} show that under constraints, $\xi_k$ corresponds to $\mathfrak{g}_k^\top\mathfrak{R}_k^{-1}\mathfrak{g}_k$, and $f_{k,i}=\bm{0}$ is the lower bound achievable only by the true value. However, it couples the unknown parameters and the quadratic terms of observation noises (when present).
The second baseline relies on equations: $\mu_{k,i}-A^\top\mu_{k+1,i}-(\begin{bmatrix}x_{k,i}^\top&1 \end{bmatrix}^\top\otimes I)^\top\bar{\mathcal{D}}_n\theta=\bm{0}$, $B^\top\mu_{k+1,i}+(u_{k,i}\otimes I)^\top\mathcal{D}_p\theta=\bm{0}$, and $\mu_{K+2,i}=\bm{0}$, where $\mu_{k,i}$ are costates. In Problem \ref{prob1}, the unknown terminal state $x_{K+1,i}$ is substituted by $Ax_{K,i}+Bu_{K,i}$ in this baseline. A linear estimation problem for $\theta$ is established similarly to \citep{d3}. Despite jointly estimating the weights and the target, it operates in open-loop and is sensitive to noise.
\subsection{Vehicle-on-a-lever Example}
\begin{figure}[!htb]
  \centering
  \includegraphics[width=1\hsize]{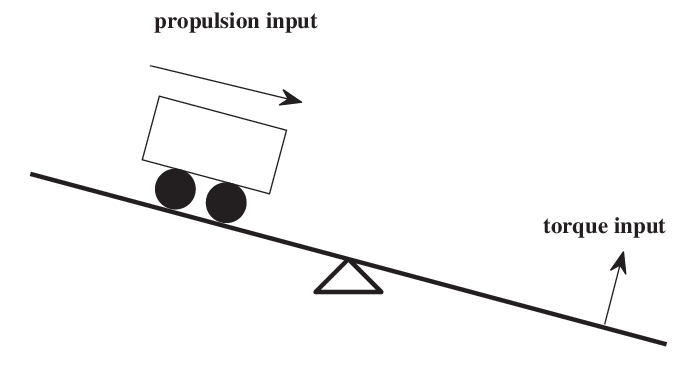}
  \caption{A vehicle, controlled by propulsion $u_1$ and torque $u_2$, moves along a uniform lever, with its center of gravity at its centroid and the lever pivoted at its center.}
  \label{fig1}
\end{figure}

\begin{figure}[!htb]
  \centering
  \includegraphics[width=1\hsize]{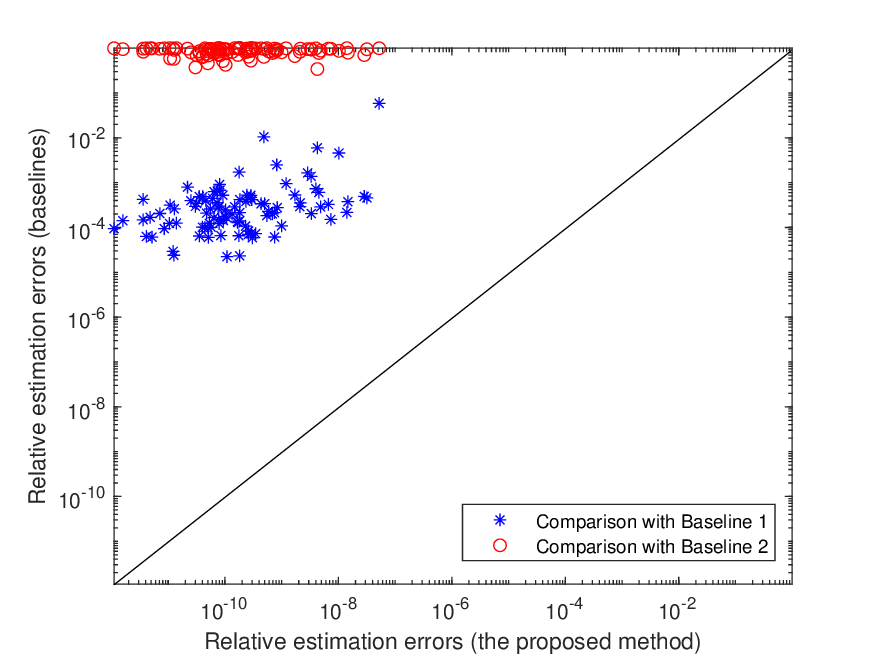}
  \caption{Relative error scatter plot for baselines and proposed method over 100 Monte Carlo trials.}
  \label{fig2}
\end{figure}

A vehicle with propulsion $u_1$ and torque $u_2$ moves along a uniform lever (Fig.~\ref{fig1}). The 4D state ${x}$ comprises the displacement from the lever center to the vehicle centroid along the lever, the lever's deviation angle from the horizontal, and their velocities. The state evolves as: $\dot{x}_1 = x_3$, $\dot{x}_2 = x_4$, $\dot{x}_3 = {u}_1/m_1 - g\sin({x}_2)$, $\dot{x}_4 = 12({u}_2 - m_1g\cos({x}_2){x}_1)/(m_2l^2)$, with $g = 9.8$. The mass $m_1 \sim \mathcal{U}[1,2]$, the length $l \sim \mathcal{U}[2,4]$, and $m_2 = 0.5l$, where $\mathcal{U}$ denotes the uniform distribution. Matrices $A$ and $B$ result from linearizing and discretizing the equations using a first-order Taylor expansion of the trigonometric functions around $0$, with a sampling period of $0.2s$. The process noise is $w_k \sim \mathcal{N}(\bm{0}, 0.1I_n)$. The observation noise is first omitted to isolate the effects of process noise, as the baselines cannot handle both noises with unknown statistics.  
The weight matrices $Q = \text{diag}(Q_0Q_0^\top, \bm{0})$ and $R = R_0R_0^\top$ are derived from $Q_0, R_0 \sim \mathcal{U}[0,1]^{2 \times 2}$. The initial state ${x_0}$ and the target $d$ position the vehicle with lever deviation within $\pm30^\circ$ of the horizontal. Simulations use $M = 5$ trajectories and $K = 8$ steps, with the sparsity pattern of $Q$ unknown. Performance is evaluated using the relative error $\|\Delta{\theta}\|/\|\theta\|$.  
Over 100 Monte Carlo trials, the proposed method achieves relative errors $<10^{-6}$, outperforming the baselines (Fig.~\ref{fig2}). Computation times: proposed $= 0.256s$, Baseline 1 (SDP) $= 34.216s$, Baseline 2 (PMP) $= 0.444s$. Besides, with added Gaussian observation noise (covariance: $10^{-4}I$), relative error mean $\pm$ standard deviation: proposed $0.1450 \pm 0.1741$, baselines $0.1953 \pm 0.2428$ and $0.7848 \pm 0.2065$.  
Under process noise alone, the proposed method offers superior efficiency compared to Baseline 1 and greater robustness than Baseline 2. When both types of noise and unknown covariances are present, all methods are unreliable with limited data; a larger dataset is considered in the next example.  
Baseline 1 uses YALMIP \citep{bx25} and MOSEK \citep{bx26} with default tolerances (e.g., \texttt{MSK\_DPAR\_INTPNT\_CO\_TOL} settings: DFEAS=$10^{-8}$, INFEAS=$10^{-12}$, MU\_RED=$10^{-8}$, NEAR\_REL=$10^{3}$, PFEAS=$10^{-8}$, REL\_GAP=$10^{-8}$). Reducing tolerances does not significantly improve Baseline 1’s relative error. Simulations are run on an Intel i7-8750H CPU @ 2.20GHz with 16GB RAM.
\subsection{Multi-agent Formation Control Example}
\begin{figure}[!htb]
  \centering
  \includegraphics[width=1\hsize]{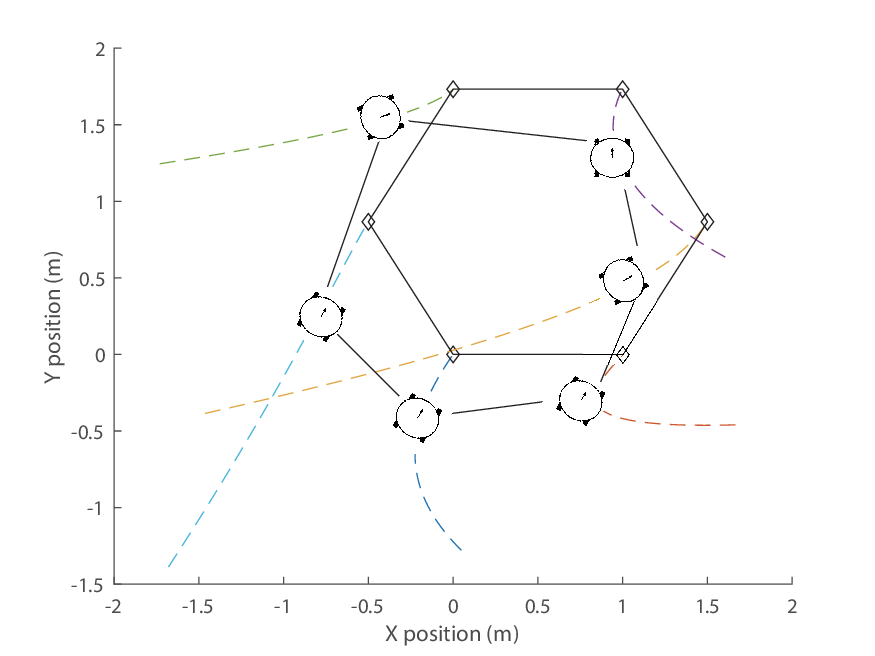}
  \caption{Trajectories of multi-agent LQT formation under fixed topology.}
  \label{fig3}
\end{figure}

\begin{figure}[!htb]
  \centering
  \includegraphics[width=1\hsize]{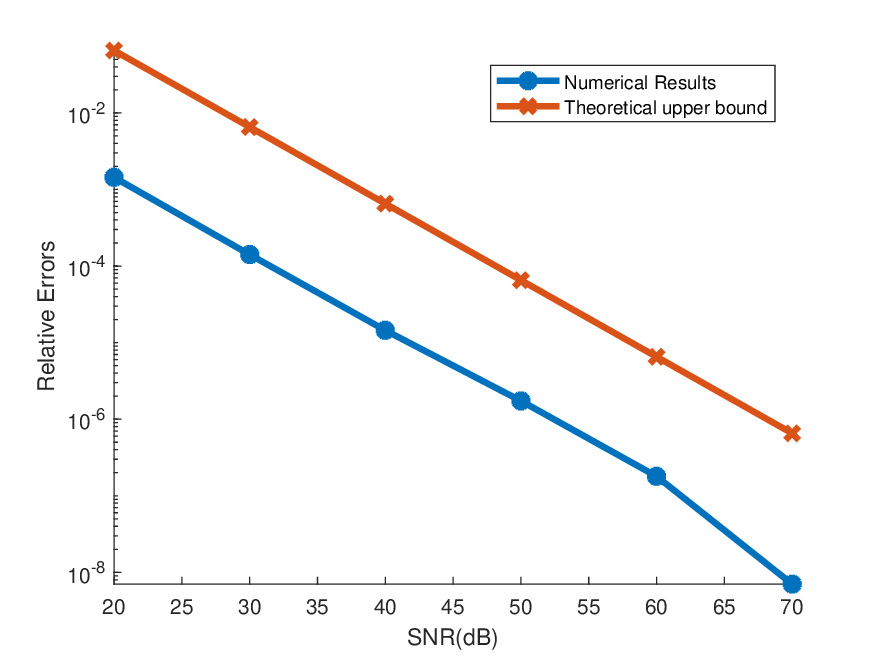}
  \caption{Relative errors versus SNR for $M = 1300$: proposed method (blue) and theoretical upper bound (red).}
  \label{fig4}
\end{figure}

\begin{figure}[!htb]
  \centering
  \includegraphics[width=1\hsize]{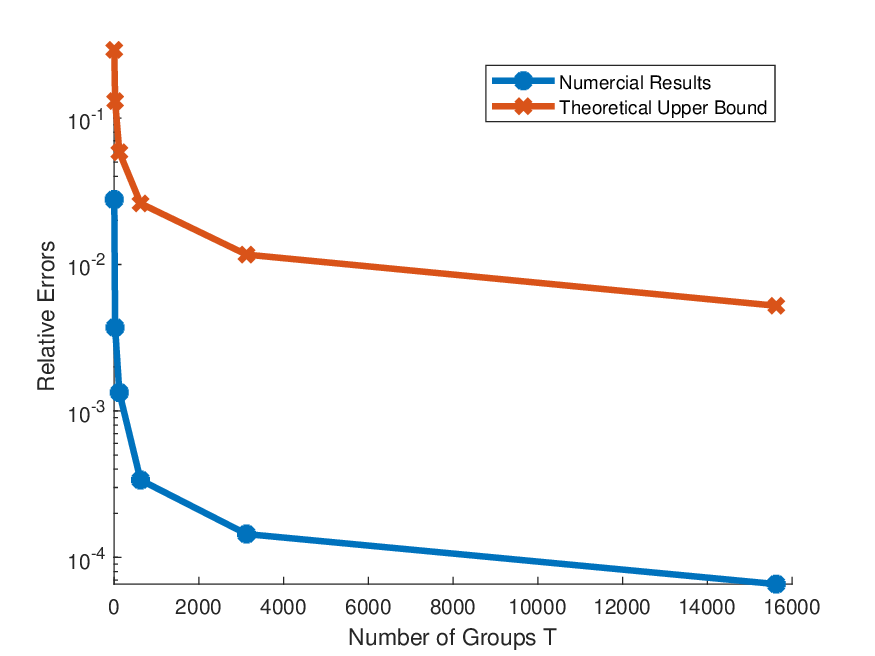}
  \caption{Relative errors versus $T$ at $\text{SNR} = 20$ dB: proposed method (blue) and theoretical bound (red).}
  \label{fig5}
\end{figure}

The multi-agent LQT formation under a fixed but unknown topology from \citep{bz3} is tested, transforming the IOC problem into identifying collective coordination patterns and intents, as described in Section~\ref{sec6}. Each agent's dynamics is $\dot{{x}}^j = {u}^j$, with a control interval of 0.01 seconds. Six agents are considered, each with 2D coordinates as states and velocities as inputs, yielding $n = p = 12$.  
The weight matrices are $Q = (L + I) \otimes I$ and $R = 0.15I$, where  
$
L = \begin{bmatrix}
3 & -1 & -1 & 0 & 0 & -1 \\
-1 & 3 & -1 & 0 & -1 & 0 \\
-1 & -1 & 3 & -1 & 0 & 0 \\
0 & 0 & -1 & 1 & 0 & 0 \\
0 & -1 & 0 & 0 & 1 & 0 \\
-1 & 0 & 0 & 0 & 0 & 1
\end{bmatrix}
$.
The target states are  
$d^1 = \begin{bmatrix} 0 & 0 \end{bmatrix}^\top$,  
$d^2 = \begin{bmatrix} 1 & 0 \end{bmatrix}^\top$,  
$d^3 = \begin{bmatrix} \frac{3}{2} & \frac{\sqrt{3}}{2} \end{bmatrix}^\top$,  
$d^4 = \begin{bmatrix} 1 & \sqrt{3} \end{bmatrix}^\top$,  
$d^5 = \begin{bmatrix} 0 & \sqrt{3} \end{bmatrix}^\top$,  
$d^6 = \begin{bmatrix} -\frac{1}{2} & \frac{\sqrt{3}}{2} \end{bmatrix}^\top$.  
Fig.~\ref{fig3} shows a typical formation trajectory, with circles representing agents, lines indicating current formations, colored dashed lines as trajectories, hollow rhombuses as endpoints, and endpoint connections forming the final formation.  
The data length is $K = 24$. Initial components are sampled uniformly from $[-2, 2]$. Process noise $\bar{w}_k$ is a zero-mean normal vector with variance 0.1 added to $x_k$.  
Without observation noise ($M = 13$), the relative error $\|\Delta{\theta}\|/\|\theta\| < 10^{-14}$, and all $d^j$ are recovered exactly.  
With observation noise ($M = 1300$, $s = 13$, $T = 100$), the relative error versus SNR is shown in Fig.~\ref{fig4}. Noise is applied to $X_k$ and $U_k$ using standard normal random variables generated by MATLAB's \texttt{randn}, scaled to achieve different SNR levels (SNR$=20:10:70$).  
Fig.~\ref{fig4} includes the theoretical bound from (\ref{sim1}). The relative error remains strictly below the bound and decreases with increasing SNR.  
For $T$-convergence ($s = n + 1 = 13$, $M = 13T$, $\text{SNR} = 20$ dB), the relative error versus $T$ is shown in Fig.~\ref{fig5} ($T=5^{1:1:6}$). The target states are identifiable; for example, at $T = 25$:  
$
|\Delta d^1| = \begin{bmatrix} 0.0040 & 0.0065 \end{bmatrix}^\top,
|\Delta d^2| = \begin{bmatrix} 0.0053 & 0.0031 \end{bmatrix}^\top, 
|\Delta d^3| = \begin{bmatrix} 0.0021 & 0.0042 \end{bmatrix}^\top,
|\Delta d^4| = \begin{bmatrix} 0.0030 & 0.0030 \end{bmatrix}^\top, 
|\Delta d^5| = \begin{bmatrix} 0.0028 & 0.0059 \end{bmatrix}^\top,
|\Delta d^6| = \begin{bmatrix} 0.0062 & 0.0025 \end{bmatrix}^\top
$.
These results indicate the recovery of physically meaningful target states regardless of coordinate origin.  
\section{Conclusions}\label{sec7}
The inverse optimal control for a finite-horizon discrete-time LQT problem has been addressed in this paper, with the aim of jointly determining the state weight matrix, the input weight matrix and the target state according to the optimal input and state trajectories. 
{A vectorization algorithm has been proposed, which significantly outperforms the methods inspired by existing works dealing with similar problems.}
In addition, theoretical guarantees have been provided for obtaining consistent estimates. Moreover, structural prior knowledge has been leveraged to reduce errors, demonstrating its potential in handling high-dimensional problems through topology identification in multi-agent systems. 

One limitation is that it requires to observe both the optimal control input and the corresponding state trajectories. Future work might extend the current results to practical scenarios with partial state observations. Besides, extending the framework to handle general input/state constraints and more complex cost structures will be a valuable direction for future research.

\begin{scriptsize}
\bibliographystyle{unsrt}
\bibliography{autosam}
\end{scriptsize}
\end{document}